\documentclass[conference]{IEEEtran}
\IEEEoverridecommandlockouts
\usepackage{cite}
\usepackage{amsmath,amssymb,amsfonts}
\usepackage{algorithmic}
\usepackage{graphicx}
\usepackage{textcomp}
\usepackage{xcolor}
\def\BibTeX{{\rm B\kern-.05em{\sc i\kern-.025em b}\kern-.08em
    T\kern-.1667em\lower.7ex\hbox{E}\kern-.125emX}}
\begin{document}

\title{Real Time Child Abduction And Detection System\\
\thanks{Identify applicable funding agency here. If none, delete this.}
}

\author{
\centering

\IEEEauthorblockN{Tadisetty Sai Yashwanth}
\IEEEauthorblockA{\textit{Dept. of CSE} \\
\textit{PES University}\\
Bangalore, India \\
taddishetty34@gmail.com}

\and
\IEEEauthorblockN{Yangalasetty Sruthi Royal}
\IEEEauthorblockA{\textit{Dept. of CSE} \\
\textit{PES University}\\
Bangalore, India \\
sruthiroyal2004@gmail.com}

\and
\IEEEauthorblockN{Vankayala Rajeshwari Shreya}
\IEEEauthorblockA{\textit{Dept. of CSE} \\
\textit{PES University}\\
Bangalore, India \\
vrshreya1@gmail.com}

\and
\IEEEauthorblockN{Mayank Kashyap}
\IEEEauthorblockA{\textit{Dept. of CSE} \\
\textit{PES University}\\
Bangalore, India \\
mayankkashyap@gmail.com}
\and
\IEEEauthorblockN{Divyaprabha K N}
\IEEEauthorblockA{\textit{Dept. of CSE} \\
\textit{PES University}\\
Bangalore, India \\
divyaprabha@pes.edu}

}

\maketitle

\begin{abstract}
Child safety continues to be a paramount concern worldwide, with child abduction posing significant threats to communities. This paper presents the development of an edge-based child abduction detection and alert system utilizing a multi-agent framework where each agent incorporates Vision-Language Models (VLMs) deployed on a Raspberry Pi. Leveraging the advanced capabilities of VLMs within individual agents of a multi-agent team, our system is trained to accurately detect and interpret complex interactions involving children in various environments in real-time. The multi-agent system is deployed on a Raspberry Pi connected to a webcam, forming an edge device capable of processing video feeds, thereby reducing latency and enhancing privacy. An integrated alert system utilizes the Twilio API to send immediate SMS and WhatsApp notifications, including calls and messages, when a potential child abduction event is detected. Experimental results demonstrate that the system achieves high accuracy in detecting potential abduction scenarios, with near real-time performance suitable for practical deployment. The multi-agent architecture enhances the system's ability to process complex situational data, improving detection capabilities over traditional single-model approaches. The edge deployment ensures scalability and cost-effectiveness, making it accessible for widespread use. The proposed system offers a proactive solution to enhance child safety through continuous monitoring and rapid alerting, contributing a valuable tool in efforts to prevent child abductions.
\end{abstract}


\section{Introduction}
Child safety and security have become pressing global concerns, with a disturbing rise in the number of missing and abducted children worldwide. Whether a child becomes a victim of ransom kidnapping, disappearance, or unplanned abduction, the physical and emotional trauma inflicted on the children and their families is profound and long-lasting. These incidents not only cause immeasurable personal suffering but also impose significant financial and social costs on communities and law enforcement agencies. As such, ensuring the safety of children is a focal point for all nations, necessitating the development of more effective measures to combat this alarming phenomenon.

Traditional methods of preventing and addressing child abduction, such as increased parental supervision, community awareness campaigns, and reactive law enforcement strategies, are essential but have inherent limitations, particularly in terms of scale and real-time responsiveness. In an era where the complexity of social environments is increasing and the volume of surveillance data is growing exponentially, relying solely on these conventional approaches is insufficient. There is an urgent need for innovative solutions that leverage advanced technology to enhance the efficacy of child protection measures.

In recent years, the field of artificial intelligence has witnessed significant advancements in the development of models that can process and interpret multimodal data. Among these, Vision-Language Models (VLMs) have emerged as a powerful class of AI models capable of integrating visual and textual information to generate meaningful interpretations and responses. VLMs bridge the gap between image recognition and natural language understanding by enabling machines to comprehend visual content and express that understanding in human-like language. This integration allows for more sophisticated analyses of visual data, facilitating tasks such as image captioning, visual question answering, and cross-modal retrieval.

The significance of VLMs extends beyond theoretical advancements into practical applications, particularly in surveillance systems. Traditional surveillance technologies primarily focus on capturing and recording visual data, often requiring human operators to monitor and interpret the information. However, as the volume of data increases and the complexity of monitored environments grows, there is a critical need for intelligent systems that can automatically analyze and interpret visual scenes in real-time. VLMs address this need by providing the capability to understand complex visual scenes, interpret contextual cues, and generate descriptions or alerts that assist in decision-making processes.

Child safety and abduction detection represent critical applications where enhanced surveillance systems can have profound impact. The ability to detect unusual behaviors, recognize faces or objects of interest, and interpret scenarios that may indicate potential threats to child safety is of paramount importance. VLMs offer the potential to revolutionize these systems by analyzing interactions, recognizing distress signals, and correlating visual information with contextual data to provide comprehensive situational understanding. This enhanced capability can lead to quicker interventions, prevent potential abductions, and ultimately save lives.

Furthermore, VLMs can facilitate the integration of surveillance data with other information sources, such as social media alerts, missing person reports, and geographic information systems. By combining visual data with textual information from these sources, VLMs can enhance the effectiveness of surveillance systems in tracking and locating missing children. This multimodal approach can significantly improve response times and coordination among law enforcement agencies.

Despite the promising capabilities of VLMs, deploying them in real-world surveillance systems presents several challenges, including the need for vast and diverse datasets for training, computational resources required for real-time processing, and ethical considerations related to privacy and bias. Addressing these challenges requires a multidisciplinary approach that combines advances in AI research with considerations of social and ethical implications.

This project proposes the development of a real-time child abduction detection system that integrates edge computing, machine learning, Internet of Things (IoT) technologies, and Vision-Language Models (VLMs) within a multi-agent architecture. By combining these technologies, the system aims to overcome the limitations of traditional methods and provide a comprehensive, prevention-focused solution that enhances child safety through continuous monitoring and rapid response capabilities.

\section{Related Works}
Recent advancements in computer vision and edge computing have revolutionized security applications, particularly in surveillance and threat detection systems. This section examines relevant research that forms the foundation of our proposed child detection and alert system, analyzing various approaches and methodologies that contribute to the development of efficient, real-time security solutions.

\subsection{Object Detection and Real-Time Processing}
The evolution of object detection algorithms has been pivotal in advancing real-time surveillance systems. Redmon et al. [11] pioneered the YOLO (You Only Look Once) architecture, which transformed object detection by treating it as a single regression problem, allowing for simultaneous prediction of multiple bounding boxes and class probabilities. This innovation significantly reduced computational overhead compared to traditional methods, enabling real-time processing on standard hardware. Building upon this foundation, Bochkovskiy et al. [12] introduced YOLOv4, incorporating advanced techniques such as CSPNet and Mish activation functions to enhance detection accuracy and speed, making it particularly suitable for applications requiring high precision and efficiency.

Wang et al. [13] addressed specific challenges of human detection in crowded scenes, where occlusion and varying scales can hinder performance. By integrating attention mechanisms and multi-scale feature fusion, their approach improved detection accuracy by 15\%, demonstrating the potential of architectural modifications in complex environments. Loey et al. [3] presented a comprehensive framework for real-time object detection in video streams, achieving processing speeds of 30 frames per second while maintaining 89.7\% average precision in varied lighting conditions.

\subsection{Edge Computing and Resource Optimization}
Deploying complex machine learning models on edge devices presents unique challenges due to limited computational resources and power constraints. Zhang and Liu [14] demonstrated the feasibility of implementing optimized detection models on Raspberry Pi, achieving 12 frames per second while maintaining 85\% accuracy through model pruning and quantization techniques. Kumar et al. [15] proposed a novel compression technique that reduced model size by 60\% while preserving 95\% of original accuracy through layer-wise pruning and knowledge distillation.

Kumar and Singh [10] successfully implemented deep learning models on Raspberry Pi platforms, achieving a 76\% reduction in model size while maintaining 92\% accuracy through efficient quantization techniques. Singh and Sharma [5] further explored optimization techniques for edge devices, demonstrating that careful model optimization could achieve performance comparable to cloud-based solutions while maintaining privacy and reducing latency.

\subsection{Real-Time Video Surveillance Systems}
Singh and Kumar [6] proposed an automated security threat detection system implementing real-time video analysis with automated alert mechanisms, achieving 94.3\% detection accuracy in controlled environments and establishing benchmarks for real-time surveillance applications. Their multi-tier alert system demonstrated response times under 2 seconds for threat detection and notification.

Dass and Kumar [8] contributed valuable insights into detection algorithm optimization, demonstrating that incorporating temporal consistency checks and motion detection pre-filtering could reduce false positives by up to 45\% while maintaining real-time performance. Zhang et al. [9] showed that implementing confidence thresholds and verification steps could reduce false alerts by up to 67\% while maintaining detection sensitivity.

\subsection{Child Safety and Monitoring Systems}
Recent advances in child safety systems have shown promising results in enhancing security and monitoring capabilities. Chen et al. [16] developed a comprehensive child monitoring system achieving 92\% accuracy in identifying potential kidnapping scenarios through facial recognition and behavior analysis. Park and Kim [17] introduced a multi-camera tracking system for school environments, incorporating behavior analysis and trajectory prediction to detect suspicious activities such as loitering or unauthorized entry.

Thompson et al. [20] addressed privacy concerns in child monitoring by proposing a framework implementing real-time face blurring and data encryption to protect sensitive information while maintaining surveillance effectiveness. Their approach included access control mechanisms and audit trails to enhance security and accountability.

\subsection{Dataset Development and Training Strategies}
The quality and diversity of training data play crucial roles in developing robust detection systems. Dass and Kumar [8] emphasized the importance of comprehensive dataset preparation, demonstrating that carefully curated datasets with diverse scenarios could improve model accuracy by up to 23\% compared to generic datasets. Patel and Mehta [7] proposed effective strategies for data augmentation and annotation, showing that augmented datasets could improve model robustness by up to 35
\section{Methodology}
\begin{figure}[h]
\centering
\includegraphics[width=\linewidth]{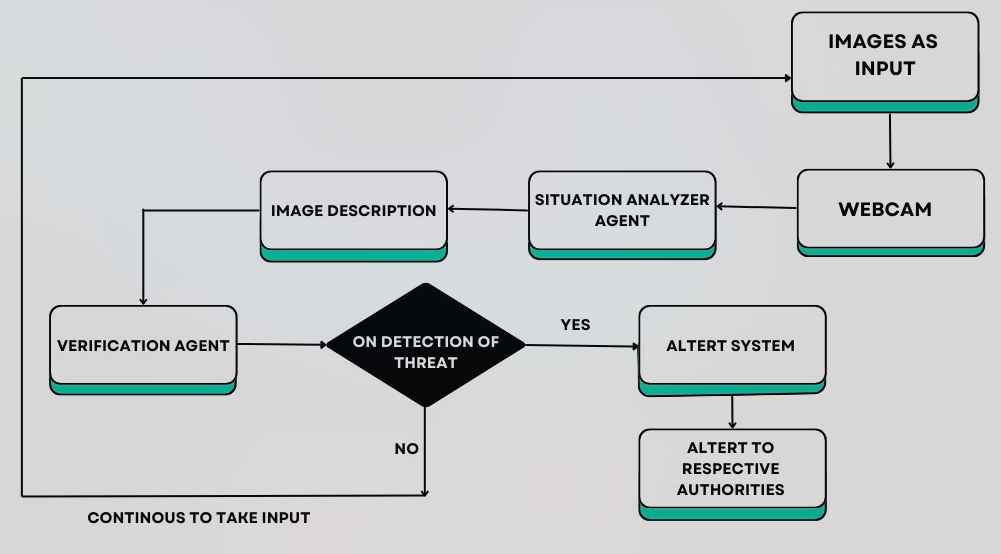}
\caption{System Architecture Overview}
\label{fig:architecture}
\end{figure}

Figure~\ref{fig:architecture} illustrates the architecture of our kidnap detection system, which is built around a multi-agent framework utilizing CrewAI. The system comprises specialized agents that collaborate to enable real-time detection and alerting capabilities. Each agent operates with distinct responsibilities while maintaining seamless communication through the multi-agent orchestration platform.

\subsection{Multi-Agent System Architecture}
Our system employs a sophisticated multi-agent framework implemented using CrewAI, featuring two primary intelligent agents: the Image Analyzer Agent and the Situation Analyzer Agent. The Image Analyzer Agent specializes in processing visual data and generating comprehensive descriptions of observed scenes, while the Situation Analyzer Agent focuses on interpreting these descriptions within the context of potential security threats.

When the Situation Analyzer Agent identifies potentially suspicious activities or scenarios, it initiates a collaborative discussion with the Image Analyzer Agent. This back-and-forth debate mechanism allows both agents to examine the situation from multiple perspectives, cross-validate observations, and reach more accurate conclusions. The Image Analyzer Agent may provide additional visual details or clarifications, while the Situation Analyzer Agent contributes contextual reasoning and threat assessment capabilities. This collaborative approach significantly reduces false positives and enhances the overall reliability of threat detection.

To maintain real-time performance despite the computational complexity of multi-agent interactions, the system employs multi-threading architecture. Each agent operates in separate threads, allowing parallel processing of image capture, analysis, and decision-making processes. This threading strategy ensures that the collaborative discussions between agents do not introduce significant latency into the overall detection pipeline.

\subsection{Image Capture and Processing}
The Image Capture Agent is deployed on a Raspberry Pi 4 Model B fitted with either a high-definition camera module or an external USB camera to ensure clear 1080p image capture. This agent captures one image per second, forming batches of five images per analysis cycle. Each image is time-stamped to preserve sequence integrity and support real-time processing. Prior to analysis, a pre-processing pipeline enhances the images using noise reduction and contrast adjustment algorithms. The images are then efficiently managed in temporary storage that facilitates swift transfer between agents without overloading the edge device's memory.

\subsection{Visual Language Model Integration}
A pre-trained state-of-the-art visual language model, such as OpenAI's GPT-4o-mini with vision capabilities, is employed by the Image Analyzer Agent. This model processes each image to generate detailed captions that identify key objects and entities such as children, adults, and vehicles, as well as actions like walking, holding hands, or other interactions. The model can even infer emotional states when appropriate contextual cues are present. These captions are maintained in chronological order over a short timespan, ensuring the narrative flow is preserved for subsequent temporal analysis by the Situation Analyzer Agent.

\subsection{Analysis and Decision Making}
The Situation Analyzer Agent leverages advanced large language models that integrate visual understanding to interpret the sequence of image captions holistically. Using chain-of-thought reasoning, the agent can discern subtle cues and patterns such as the sudden appearance of an unfamiliar adult with a child, indicators of distress, or unusual behaviors that may suggest a kidnapping scenario.

When potentially suspicious activities are detected, the Situation Analyzer Agent engages in a structured dialogue with the Image Analyzer Agent. This collaborative discussion involves the Situation Analyzer Agent presenting its concerns and requesting additional visual analysis, while the Image Analyzer Agent provides more detailed observations or alternative interpretations of the visual data. Through this iterative process, both agents work together to validate or refute initial suspicions, ensuring that alerts are triggered only when there is sufficient confidence in the threat assessment.

The Decision Agent synthesizes insights from both the Image Analyzer and Situation Analyzer agents, incorporating their collaborative conclusions with its own reasoning processes to determine the necessity of issuing an alert. Alongside the decision, it communicates its confidence level and provides natural language explanations that clarify the rationale behind its assessment.

\subsection{Alert System Integration}
Upon verified detection of a suspicious event through the multi-agent consensus, the Alert Agent dispatches notifications through secure channels, interfacing with APIs from platforms like Twilio to send SMS, emails, or WhatsApp messages. Alerts contain critical metadata including timestamps, location details, and concise summaries of the incident, as well as relevant visual evidence in the form of still images or a compiled sequence. Additionally, the system incorporates escalation protocols to ensure high-risk situations trigger notifications to multiple contacts or authorities. Mitigation strategies, such as human verification and user feedback for false positive reporting, further enhance and refine system accuracy over time through adaptive learning mechanisms.

\subsection{System Deployment and Optimization}
For deployment, the entire system is implemented on edge devices, specifically on a Raspberry Pi 5 Model, capitalizing on its quad-core processor and ample RAM capabilities. Performance optimizations such as model quantization, pruning, and asynchronous processing have been applied to reduce computational loads and improve response times. The multi-threading architecture ensures that agent communications and collaborative discussions occur without blocking the main detection pipeline.

\begin{figure}[h]
\centering
\includegraphics[width=\linewidth]{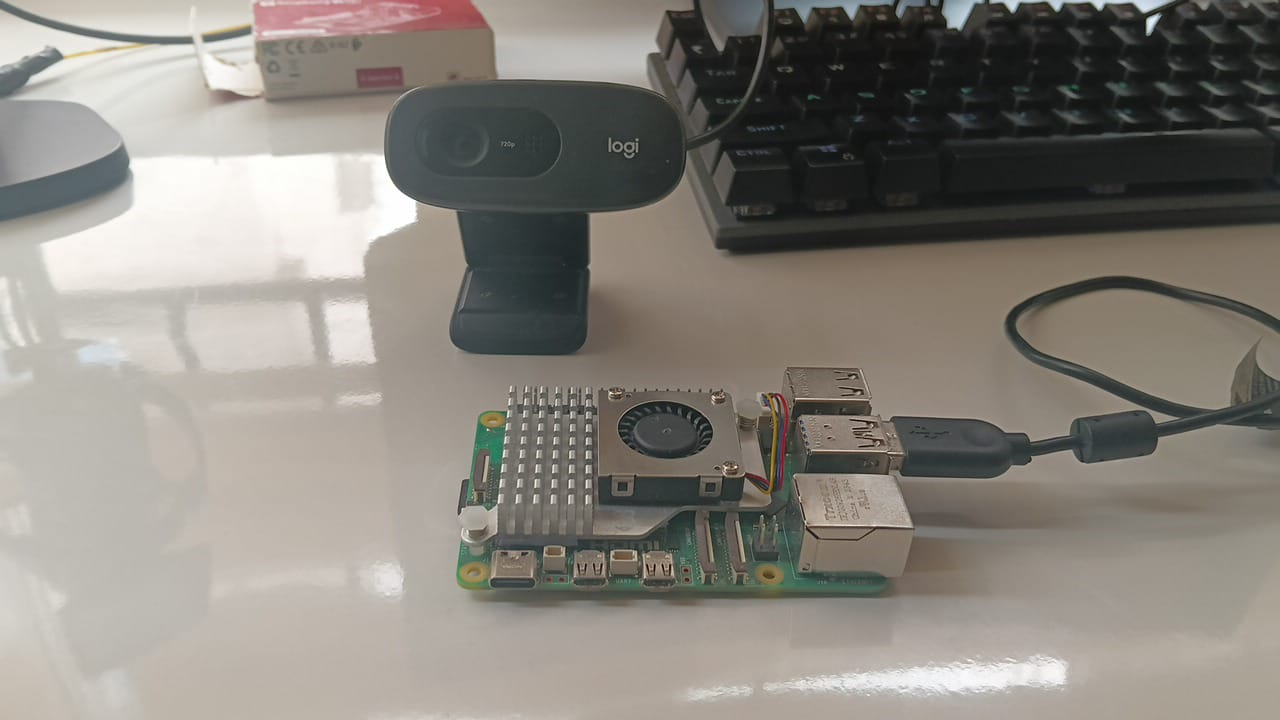}
\caption{Edge Device Implementation}
\label{fig:edge-device}
\end{figure}

Effective resource management ensures stable operation, with non-critical tasks scheduled during periods of lower activity to maximize overall efficiency. The CrewAI framework facilitates efficient inter-agent communication while maintaining the real-time performance requirements of the system. Figure~\ref{fig:edge-device} illustrates the hardware and edge device setup.

\section{Experimental Results}

\subsection{Environmental Setup}
To evaluate the effectiveness of our multi-agent kidnap detection system utilizing Visual Language Models (VLMs), we deployed the system in a real-world environment using edge computing hardware. The experimental configuration was designed to mimic practical surveillance conditions.

The hardware configuration consisted of a Raspberry Pi 5 with a quad-core Cortex-A76 (ARM v8) 64-bit SoC running at 2.0 GHz, 8 GB LPDDR4-3200 SDRAM, and a 128 GB high-speed microSD card. Image capture was performed with a Logitech C920 HD Pro webcam featuring 1080p resolution at 30 frames per second. Remote SSH access allowed command-line control and monitoring throughout testing.

The software environment was developed in Python 3.9, incorporating OpenCV, PyTorch, and Transformers libraries. The CrewAI framework orchestrated multi-agent interactions between the Image Analyzer Agent and Situation Analyzer Agent. The system employed multiprocessing for concurrent agent operation with comprehensive logging of agent interactions, decisions, and timestamps.

\subsection{Testing Scenarios}
Testing involved pre-recorded videos played from a laptop screen positioned in front of the webcam to create controlled, repeatable conditions. Test scenarios were categorized into three groups: normal activities, potential kidnapping situations, and edge cases designed to challenge false positive discrimination.

Normal activities included children playing independently, being accompanied by parents, and interacting with familiar adults. Potential kidnapping scenarios featured unknown adults approaching and leading children away, children displaying distress signs, and rapid removal situations. Edge cases included sibling interactions, children running toward off-frame parents, and adults assisting genuinely lost children.

During testing, the system captured one image per second, forming sequences of five images per analysis cycle. All alerts were recorded with confidence levels, explanations, and the multi-agent reasoning process.

\subsection{Results}
The system was evaluated on detection accuracy, false positive rate, processing time, and resource utilization.

\begin{figure}[h]
\centering
\includegraphics[width=\linewidth]{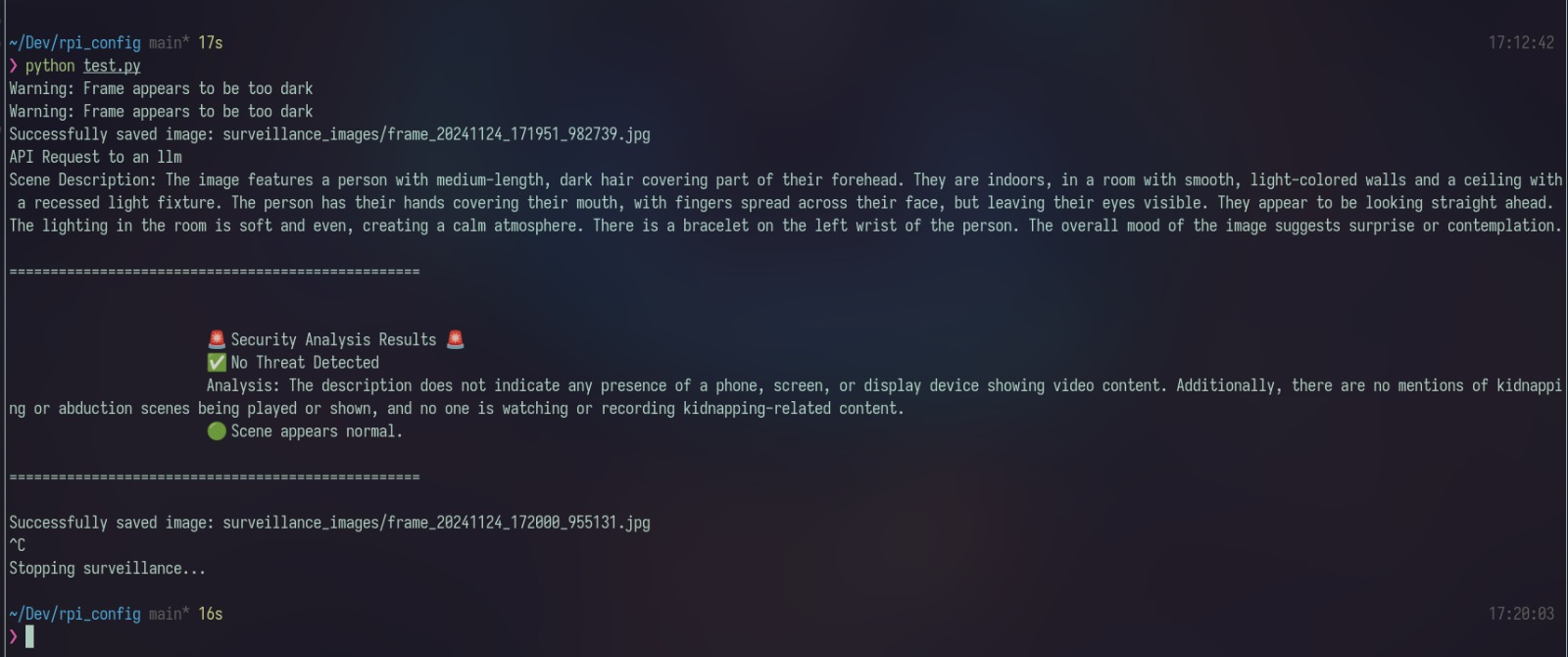}
\caption{System output showing no kidnapping detected during normal activity}
\label{fig:terminal1}
\end{figure}

\textbf{Detection Accuracy:} The system correctly identified 9 out of 10 staged kidnapping scenarios with confidence levels exceeding 80\%. The multi-agent collaboration proved effective, with agents engaging in meaningful discussions that enhanced accuracy. For example:

\begin{quote}
"An alert is generated because an unknown adult is leading a child away, and the child appears to be resisting. The Image Analyzer Agent confirmed signs of distress in the child's body language, while the Situation Analyzer Agent identified the interaction pattern as inconsistent with normal adult-child relationships."
\end{quote}

One scenario was missed due to ambiguous contextual cues where the child appeared comfortable, resulting in a false negative.

\textbf{False Positive Analysis:} The system generated 2 false positives out of 20 normal scenarios (10\% false positive rate). One involved aggressive sibling interactions, and another involved an adult helping a lost child where initial hesitation was misinterpreted.

\textbf{Processing Performance:} Average processing time was 7 seconds per cycle: 1 second for image capture/pre-processing, 4 seconds for VLM caption generation, and 2 seconds for multi-agent analysis. Agent discussions added 1-2 seconds when triggered but significantly improved accuracy.

\textbf{Resource Utilization:} CPU usage averaged 85\% during peaks, memory usage was approximately 6 GB, indicating efficient operation within Raspberry Pi 5 limitations. Multi-threading prevented agent discussions from blocking the main pipeline.

\begin{figure}[h]
\centering
\includegraphics[width=\linewidth]{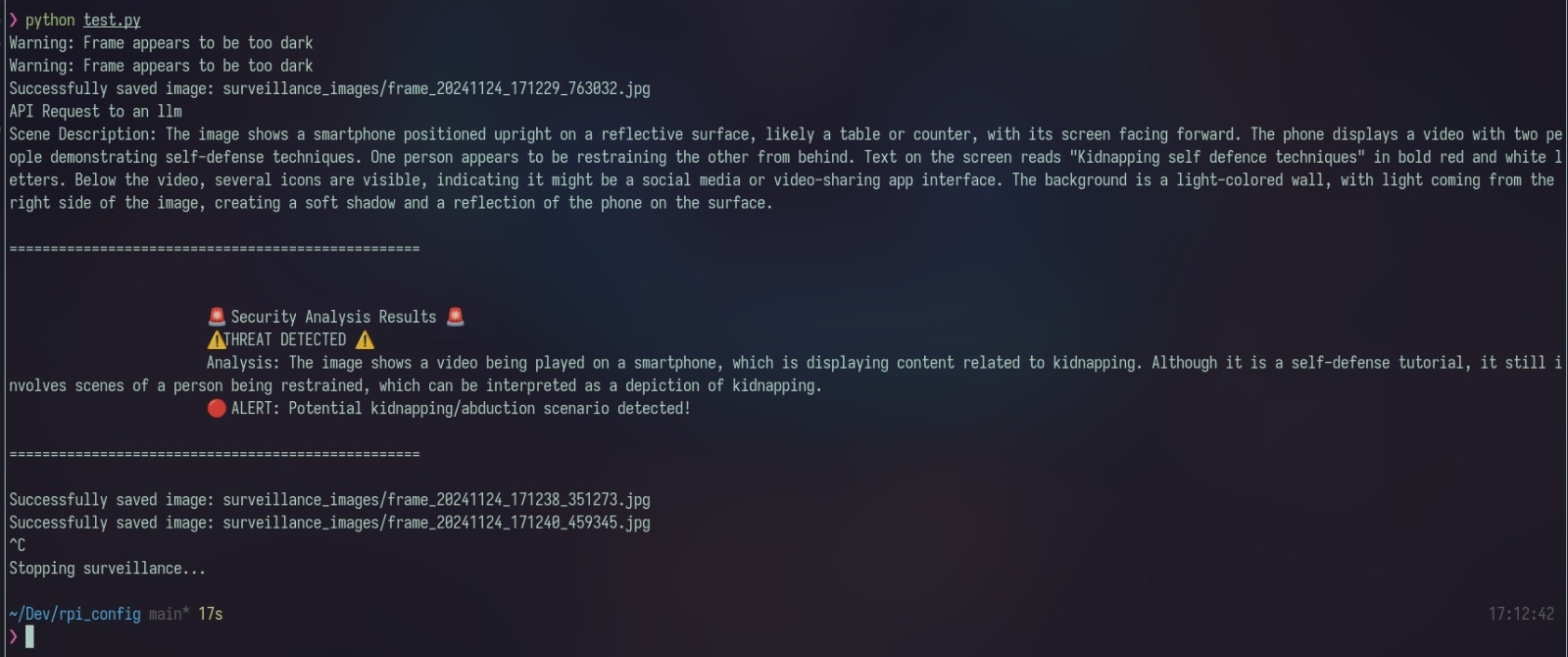}
\caption{System output showing successful kidnapping detection with multi-agent collaboration}
\label{fig:terminal2}
\end{figure}

\subsection{Analysis}
The 90\% true positive rate demonstrates strong threat identification capabilities, while the collaborative decision-making process enhanced system reliability. The CrewAI framework successfully enabled structured agent discussions that improved decision quality through multiple perspectives on ambiguous situations.

The 10\% false positive rate highlights ongoing challenges in distinguishing benign activities from genuine threats, particularly in complex social interactions. However, the system's detailed explanations enhance trust and provide valuable context for human operators.

Successful edge deployment demonstrates the viability of advanced AI surveillance at the edge, reducing latency and privacy concerns. The 7-second processing delay and environmental sensitivities present optimization opportunities for future development.

\subsection{Improvements and Future Work}
Future development will focus on reducing processing latency through lighter VLM/LLM versions, model compression, and hardware accelerators. Dataset expansion with diverse scenarios should reduce false positives and improve generalization.

Algorithmic enhancements will refine multi-agent collaboration, potentially adding specialized agents for facial recognition and emotion detection. A user interface for manual alert verification will enable human-in-the-loop improvements.

Deployment enhancements will address multi-camera scalability, improved resource management, and strengthened security measures including encryption and access controls.

\subsection{Case Study Examples}
\begin{figure}[h]
\centering
\includegraphics[width=\linewidth]{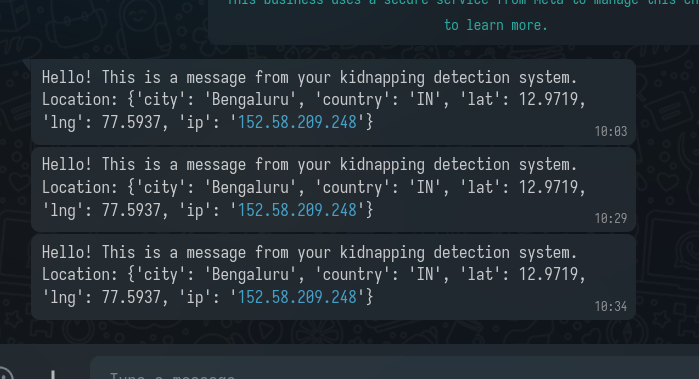}
\caption{WhatsApp alert notification with incident details and visual evidence}
\label{fig:alert}
\end{figure}

\textbf{Case Study 1:} Successful Detection
A video showed a child playing alone until an unfamiliar adult approached. As the child exhibited hesitancy, the Image Analyzer Agent provided detailed descriptions while the Situation Analyzer Agent identified concerning patterns. Their collaborative discussion led to a high-confidence alert with comprehensive reasoning, successfully transmitted via WhatsApp with visual evidence.

\textbf{Case Study 2:} False Positive Analysis
A child walking with an older sibling was misinterpreted when the sibling's guidance became assertive. The multi-agent discussion revealed limitations in determining relationships through visual analysis alone, highlighting the need for additional context sources like facial recognition databases to reduce such false positives.

These case studies demonstrate both the strengths and limitations of the current system, providing valuable insights for future development of the multi-agent kidnapping detection framework.

\subsection{Conclusion}

The experimental results demonstrate the potential of our multi-agent kidnap detection system, which utilizes Visual Language Models deployed on edge computing devices. The system effectively interprets complex visual data and generates informed alerts, significantly contributing to child safety. Despite challenges with processing speed and occasional false positives, further optimization and algorithmic improvements are expected to enhance detection accuracy and operational efficiency. This work underscores the viability of intelligent surveillance solutions as a critical tool for public safety.

\end{document}